# Imaging Properties of Two-Dimensional Microlenses

Vera N. Smolyaninova [1], Igor I. Smolyaninov [2], Alexander V. Kildishev [3], and Vladimir M. Shalaev [3]

[1] *Department of Physics Astronomy and Geosciences, Towson University, 8000 York Rd., Towson, MD 21252, USA*

[2] *Department of Electrical and Computer Engineering, University of Maryland, College Park, MD 20742, USA*

[3] *Shool of Electrical and Computer Engineering and Birck Nanotechnology Center, Purdue University, West Lafayette, IN 47907, USA*

**Despite strong experimental and theoretical evidence supporting superresolution imaging based on microlenses, imaging mechanisms involved are not well understood. Based on the transformation optics approach, we demonstrate that microlenses may act as two-dimensional fisheye or Eaton lenses. An asymetric Eaton lens may exhibit considerable image magnification, which has been confirmed experimentally.**

Current interest in electromagnetic metamaterials has been motivated by recent work on superlenses, cloaking and transformation optics [1-3]. This interest has been followed by considerable efforts aimed at introduction of metamaterial structures that could be realized experimentally. Unfortunately, it appears difficult to develop metamaterials with low-loss, broadband performance. The difficulties are especially severe in the visible frequency range where good magnetic performance is limited. On the other hand, very recently we have demonstrated that many transformation optics and metamaterial-based devices requiring anisotropic dielectric permittivity and magnetic permeability could be emulated by specially designed tapered waveguides. This approach leads to low-loss, broadband performance in the visible frequency range, which is difficult to achieve by other means. We have applied this technique to broadband electromagnetic cloaking in the visible range [4]. In this paper we apply the same technique to experimental realization of the fisheye and Eaton microlenses, which were suggested to act as ideal imaging devices even in the absence of negative refraction [5]. Realization of these microlenses using electromagnetic metamaterials would require sophisticated nanofabrication techniques. In contrast, our approach leads to a much simpler design, which involves two-dimensional (2D) imaging using a small liquid microdroplet.

Despite strong experimental and theoretical evidence supporting superresolution imaging based on microlenses and microdroplets, imaging mechanisms involved are not well understood. Imaging by surface plasmon polaritons (SPP) [6] has been proposed as the main super-resolution mechanism in imaging experiments using glycerin microdroplets on gold film surface [7]. Resolution of the order of $\lambda/8$ has been observed in these experiments. On the other hand, magnification of near-field image components has been suggested in recent experiments with self-assembled plano-spherical nanolenses [8], which demonstrated resolution of the order of $\lambda/4$. Our analysis in terms of the effective metamaterial parameters indicates that the shape of microlenses and microdroplets provides natural realization of the effective refractive index distribution in the fisheye and Eaton microlenses [4]. The starting point of our analyses is the dispersion law of guided modes in a tapered waveguide. In case of the metal-coated dielectric waveguide it can be written in a simple analytical form:

$$\frac{\omega^2 n_d}{c^2} = k_x^2 + k_y^2 + \frac{\pi^2 l^2}{d(r)^2} \quad (1)$$

where $n_d$ is the refractive index of the dielectric, $d(r)$ is the waveguide thickness, and $l$ is the transverse mode number. We assume that the thickness $d$ of the waveguide in the $z$-direction changes adiabatically with radius $r$. A photon launched into the $l$-th mode of the waveguide stays in this mode as long as $d$ changes adiabatically [9]. If we wish to emulate refractive index distribution $n(r)$ of either 2D fisheye or 2D Eaton lens:

$$\frac{\omega^2 n(r)}{c^2} = k_x^2 + k_y^2 \quad (2)$$

we need to produce the following profile of the microdroplet:

$$d = \frac{l\lambda}{2\sqrt{n_d^2 - n^2(r)}} \quad (3)$$

This is easy to do for some particular mode $l$ of the waveguide. Typical microdroplet/microlens profiles which emulate fisheye lens:

$$n = \frac{2n_1}{1 + \frac{r^2}{R^2}} \quad (4)$$

or Eaton lens:

$$n = 1 \text{ for } r < R, \text{ and } n = \sqrt{\frac{2R}{r} - 1} \text{ for } r > R \quad (5)$$

are shown in Fig. 1. Real glycerin microdroplets have shapes, which are somewhere in between these cases. Since the refractive index distribution in the fisheye lens is obtained via the stereographic projection of sphere onto a plane [5], points near the droplet edge correspond to points located near the equator of the sphere. Therefore, these points are imaged into points located near the opposite droplet edge, as shown in Fig.2(a). The Eaton lens has similar imaging properties, as shown in Fig.2(b). We have tested this imaging mechanism using glycerin microdroplets formed on the surface of gold film, which were illuminated near the edge using tapered fiber tips of a near-field scanning optical microscope (NSOM), as shown in Fig.3. As expected from the numerical simulations, an image of the NSOM tip was easy to observe at the opposite edge of the microdroplet.

While the fisheye lens design is difficult to modify to achieve image magnification, modification of the Eaton lens is straightforward. As shown in Fig.4, two halves of the Eaton lens having different values of parameter R can be brought together to achieve image magnification. The image magnification in this case is $M=R_1/R_2$. Our numerical simulations in the case of $M=2$ are presented. Since the sides of the lens play no role in imaging, the overall shape of the imaging device can be altered to achieve the shape of a "deformed droplet". Using experimental technique described in the "Methods" section, we have created glycerin droplets with shapes, which are very close to the shape of the "deformed droplet" used in the numerical simulations. Image magnification of the "deformed droplet" has been tested by moving the NSOM probe tip along the droplet edge, as shown in Fig.5. It appears to be close to the $M=2$ value predicted by the simulations (see "Methods").

In conclusion, we have demonstrated that small dielectric microlenses behave as two-dimensional imaging devices, which can be approximated by 2D fisheye and/or Eaton lenses. Superresolution imaging observed in these microlenses is consistent with the transformation optics mechanism described in ref.[5]. In addition, deformed microlenses/microdroplets were observed to exhibit image magnification, which is consistent with numerical predictions.

**METHODS**

In our imaging experiments the "deformed droplets" were formed in desired locations by bringing a small probe Fig.6(a) wetted in glycerin into close proximity to the sample surface. The probe was prepared from a tapered optical fiber, which has an epoxy microdroplet near its apex. Bringing the probe to a surface region covered with glycerin led to a glycerin microdroplet formation under the probe (Fig.6b). The shape of the glycerin droplet was determined by the shape of the seed droplet of epoxy. The glycerin droplet under the probe can be moved to a desired location under the visual control, using a regular microscope. Our droplet deposition procedure allowed us to form droplet shapes, which were reasonably close to the shape of a magnifying Eaton lens, as shown in Fig.5. In addition, the liquid droplet boundary may be expected to be rather smooth because of the surface tension, which is essential for the proper performance of the droplet boundary as a 2D fisheye or Eaton lens.

Image magnification of the 2D magnifying Eaton lens has been measured as demonstrated in Fig.7. Position of the NSOM tip and its image in the second frame is shown by red dots in the first frame. The ratio of the gray line lengths, which connect NSOM tip and image locations in the two frames shown is close to the theoretically predicted value *M=2*.

**Figure Captions**

**Fig.1.** Typical profiles of a microdroplet which emulates either the fisheye lens ($R=7\mu m$) or the Eaton lens ($R=5\mu m$) for $l=1$.

**Fig.2.** Numerical simulations of imaging properties of the fisheye and Eaton lenses. Points near the edge of the fisheye and Eaton lenses are imaged into opposite points. Refractive index distribution in these lenses is shown in the bottom panels.

**Fig.3.** Experimental testing of the imaging mechanism of the glycerin microdroplets. The droplet is illuminated near the edge with a tapered fiber tip of a near-field scanning optical microscope (NSOM). Image of the NSOM tip is clearly seen at the opposite edge of the droplet.

**Fig.4.** Numerical simulations of image magnification ($M=2$) using the Eaton lens. Since the sides of the lens play no role in imaging, the overall shape of the imaging device can be altered to achieve the shape of a "deformed droplet".

**Fig.5.** Experimental testing of image magnification of the "deformed droplet". The NSOM probe tip was moved along the droplet edge. Bottom row presents results of our numerical simulations in the case of one and two point sources. The shape of the "deformed droplet" used in numerical simulations closely resembles the shape of the actual droplet.

**Fig.6.** The "deformed" glycerin droplets were formed in desired locations by bringing a small probe (a) wetted in glycerin into close proximity to a sample. The probe was prepared from a tapered optical fiber, which has an epoxy microdroplet near its apex. Bringing the probe to a surface region covered with glycerin led to a glycerin microdroplet formation (b) under the probe in locations indicated by the arrows.

**Fig.7.** Measurements of image magnification by the "deformed droplet": position of the NSOM tip and its image in the second frame is shown by red dots in the first frame.

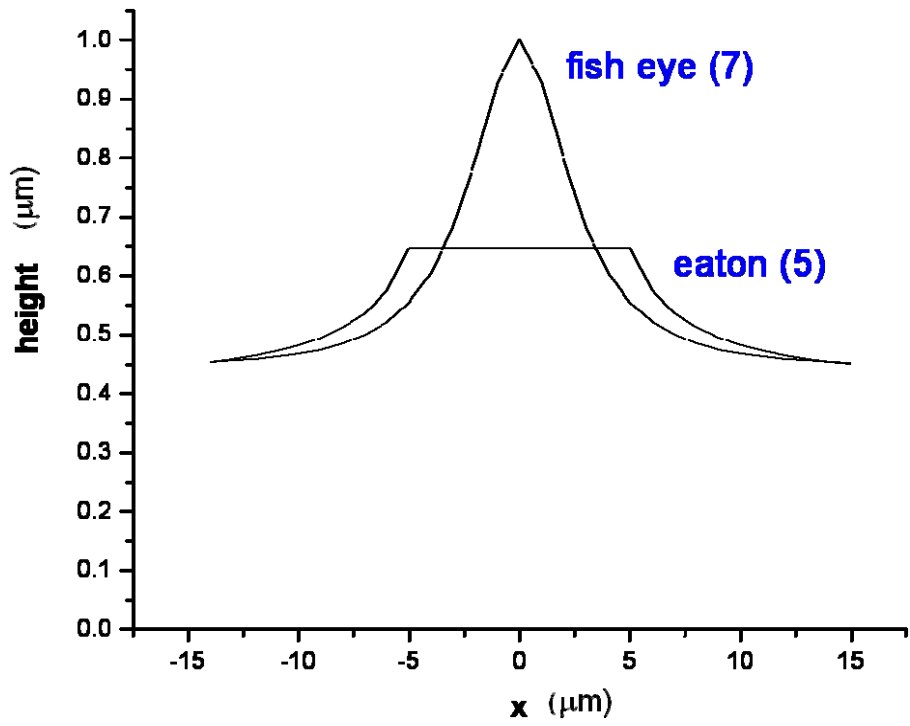

Fig. 1

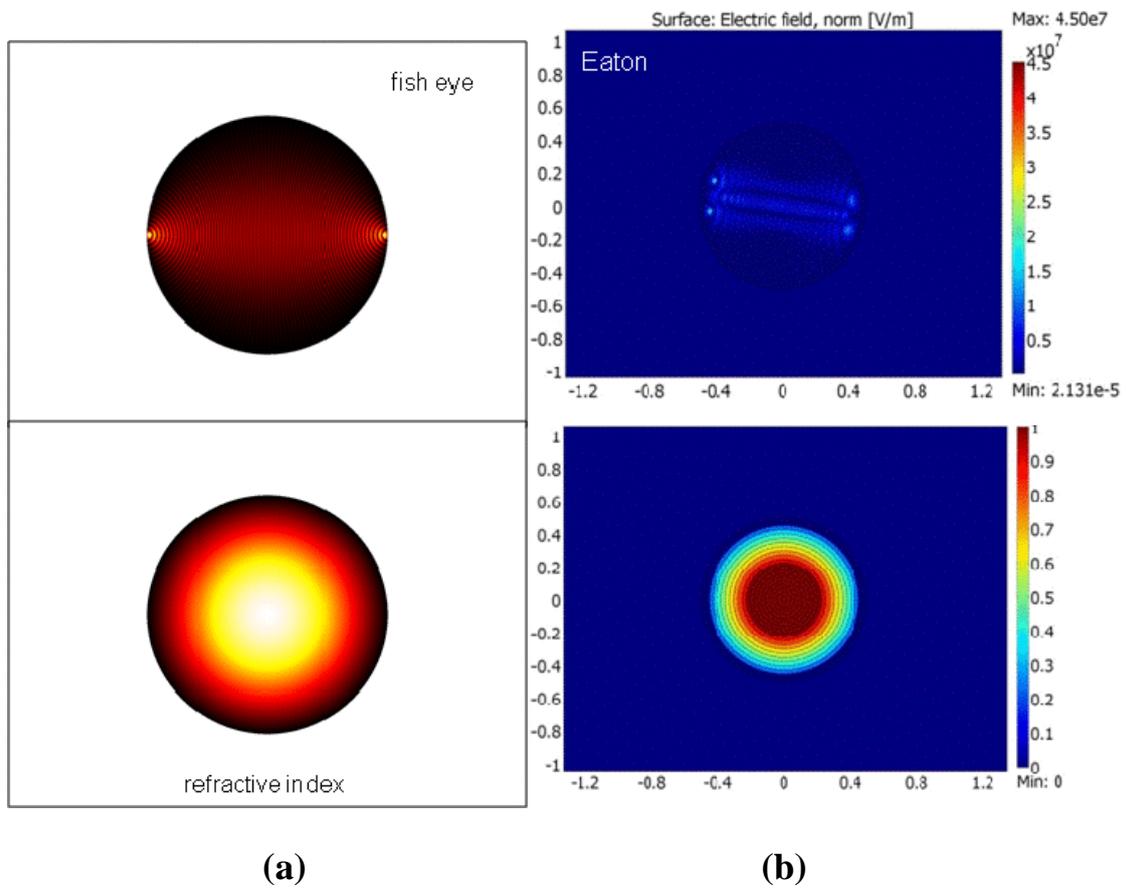

      **(a)**        **(b)**

Fig.2

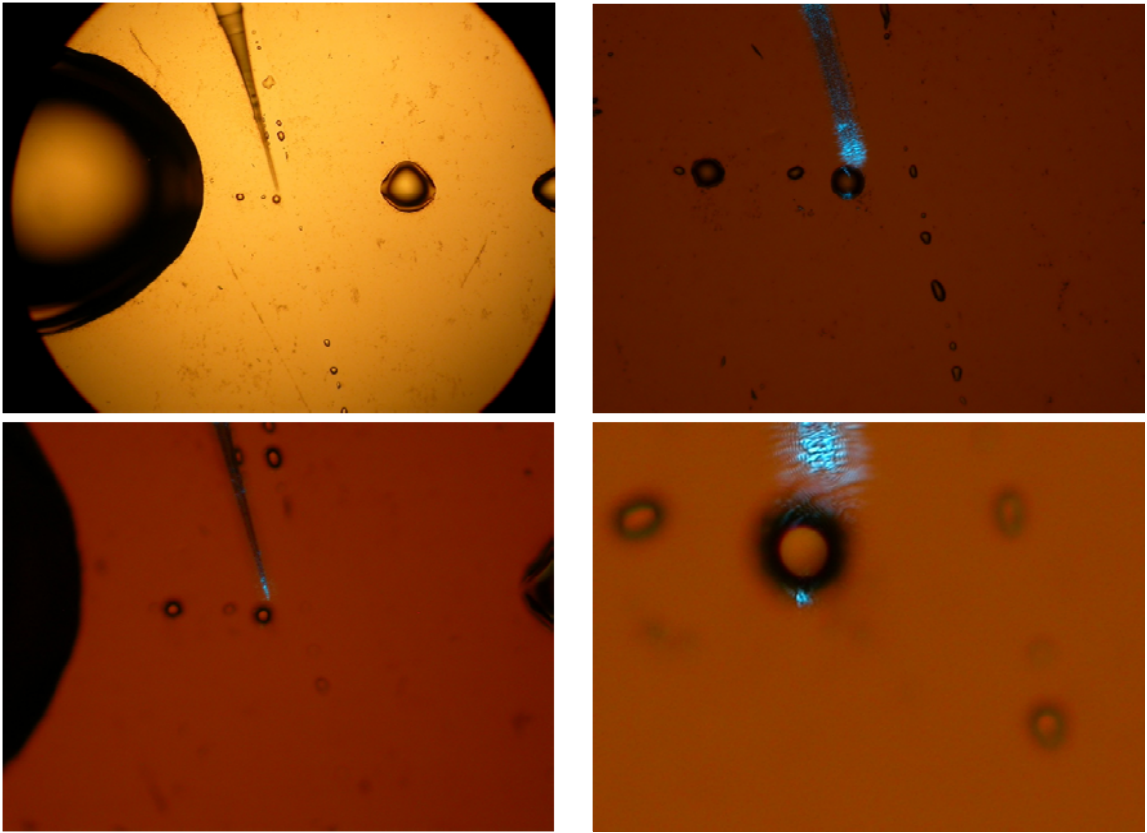

Fig.3

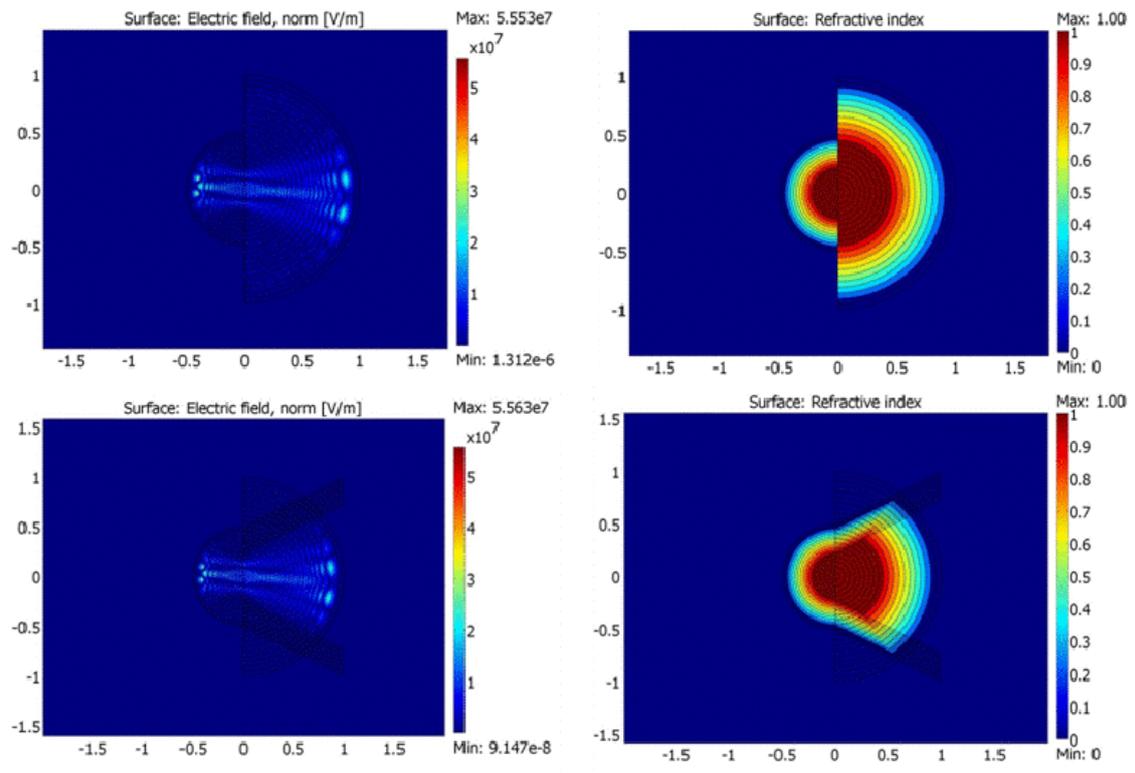

Fig.4

moving source (NSOM fiber tip) next to the "equator" as proof of magnification

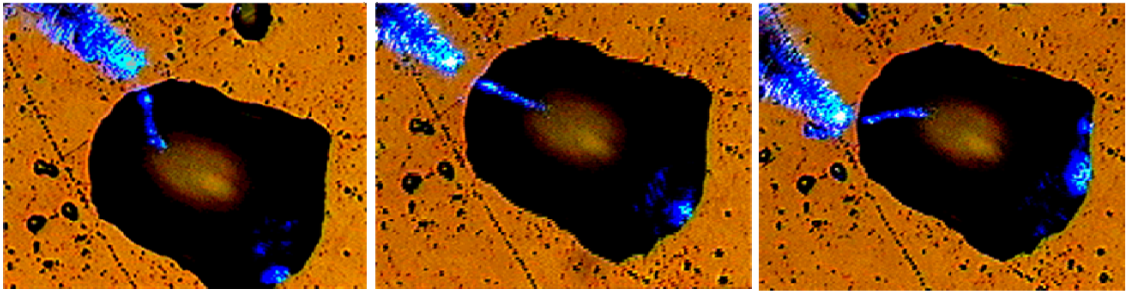

Magnification with Eaton lens

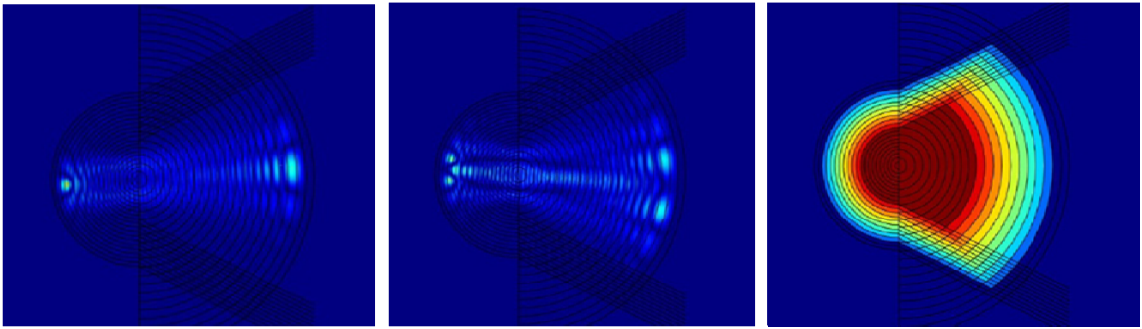

Fig.5

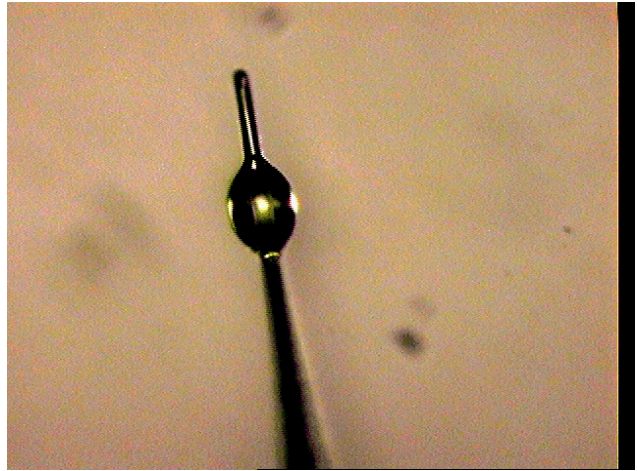

( a )

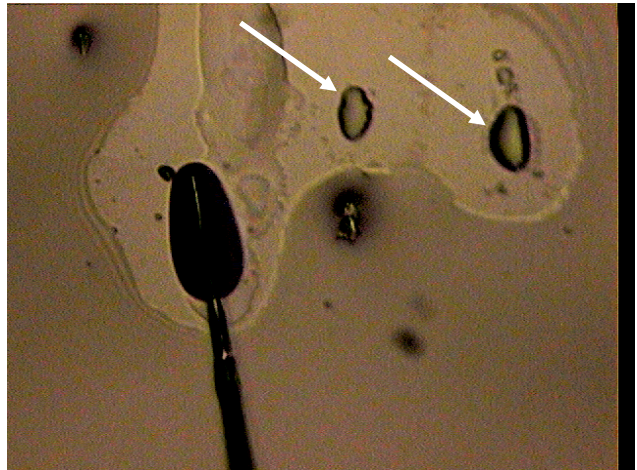

( b )

Fig.6

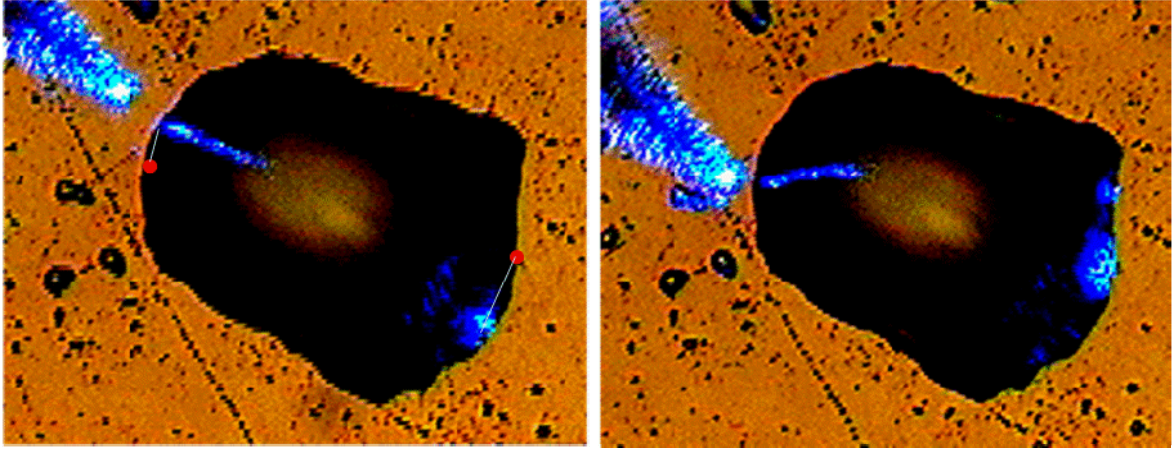

Fig.7